\documentclass[11pt,letterpaper]{article}

\usepackage[margin=1in]{geometry}
\usepackage[T1]{fontenc}
\usepackage[utf8]{inputenc}
\usepackage{lmodern}
\usepackage{amsmath,amssymb}
\usepackage{booktabs}
\usepackage{graphicx}
\usepackage{subcaption}
\usepackage{caption}
\usepackage{cite}
\usepackage[hidelinks]{hyperref}
\usepackage{microtype}
\usepackage{parskip}

\title{\textbf{Estimated Demand for Mega-Constellation Internet Service}}

\author{Akhil Rao\thanks{Rational Futures, Washington, D.C. This analysis was
conducted by the author while at NASA's Office of Technology, Policy, and
Strategy. It is a work of the U.S. Government, released under FOIA request
No.~25-00868-F-HQ, and is in the public domain. The views expressed are the
author's own and do not necessarily reflect those of NASA or the U.S.
Government.}}

\date{March 2024}

\begin{document}

\maketitle

\begin{abstract}
\noindent The near-term growth of the commercial space economy and
sustainability of the low-Earth orbit (LEO) environment depends on the
commercial prospects of large LEO satellite telecommunications constellations
(``mega-constellations''). If successful, mega-constellations can spur
competition between launch providers, which may lead to lower launch prices.
They may also demand orbital sustainability services and/or generate additional
collision risk, which may in turn affect space sustainability interests.
Consumer demand for mega-constellations' internet services is critical to their
commercial success. This analysis presents a method to estimate consumer demand
for mega-constellations' internet services, using SpaceX's Starlink over
December~2021--November~2023 as an example. This analysis finds that Starlink
users are highly concentrated in the wealthiest countries. It also finds that
consumer demand for Starlink is growing more slowly than orbital capacity is
being added and more slowly than demand for non-Starlink internet overall.
Continuation of these trends may indicate that the market for
mega-constellation internet is less lucrative than previous forecasts have
indicated.
\end{abstract}

\section*{Highlights}
\begin{itemize}
\item 73\% of estimated Starlink demand is concentrated in just five countries.
\item 66\% of estimated Starlink demand is in the wealthiest quartile of
countries.
\item A 1\% increase in non-Starlink internet traffic is associated with a
0.65\% increase in Starlink subscribers.
\end{itemize}

\section{Key Findings}

As of November~2023, this analysis showed roughly 66\% of Starlink's consumer
demand came from high-income countries (upper 25th percentile of GDP per
capita). Table~\ref{tab:income} shows the breakdown of Starlink consumer share
by GDP per capita quartiles. The top ten countries by estimated demand
comprised roughly 87\% of the consumer base, and the top five countries by
estimated demand comprised roughly 73\% of the consumer base. The United States
and Canada accounted for about 42\% and 14\% of demand, respectively.

\begin{table}[htbp]
\centering
\caption{Starlink consumer base share by GDP per capita group.}
\label{tab:income}
\begin{tabular}{@{}lrr@{}}
\toprule
\textbf{Country income} & \textbf{Mean GDP per capita} & \textbf{Consumer base} \\
                        & \textbf{(\$/person/year)}    & \textbf{share (\%)} \\
\midrule
Low income          & 4,521  & 8  \\
Lower-middle income & 15,746 & 16 \\
Upper-middle income & 34,784 & 10 \\
High income         & 66,805 & 66 \\
\bottomrule
\end{tabular}
\end{table}

Statistical analysis showed that, on average, a 1\% increase in Starlink's
available capacity was associated with a 0.58\% ($\pm$0.23\%) increase in
monthly subscribers. The same analysis also showed that, on average, a 1\%
increase in monthly internet traffic from non-Starlink sources is associated
with a 0.65\% ($\pm$0.17\%) increase in Starlink subscribers.

These findings indicate that consumer demand for satellite internet is growing
more slowly than satellite internet capacity is being added to orbit, and that
consumer demand for satellite internet is growing more slowly than demand for
non-satellite internet. If these trends continue, satellite internet providers
may be forced to cut prices in consumer markets to gain customers, or may need
to find non-consumer markets in which to sell their excess capacity.
Continuation of these trends may indicate that the market for
mega-constellation services is less lucrative than previous forecasts have
indicated, which may limit the number and size of mega-constellations that
commercial markets can sustain.

\section{Methodology}

This analysis used Domain Name System (DNS) traffic data and publicly reported
subscriber counts to construct monthly country-level timeseries of Starlink
subscribers over December~2021--November~2023. DNS acts as a ``phonebook'' for
the internet: when a website is accessed using a human-readable name like
``nasa.gov'', DNS translates the human-readable name to an Internet Protocol
(IP) address (e.g., ``192.168.1.1'') indicating the location where the content
is hosted. DNS traffic data was obtained from
Cloudflare~\cite{cloudflareradar}, the company with the largest
market share among DNS service providers. Timeseries data for normalized weekly
traffic flows are obtained from the Cloudflare Radar NetFlows API, while
country-level cross-sectional data for traffic shares over
November~2022--December~2023 were obtained from the Cloudflare Radar web
dashboard. Publicly announced Starlink subscriber counts were obtained from
Wikipedia.

\begin{figure}[htbp]
\centering
\begin{subfigure}[t]{0.48\textwidth}
\centering
\includegraphics[width=\textwidth]{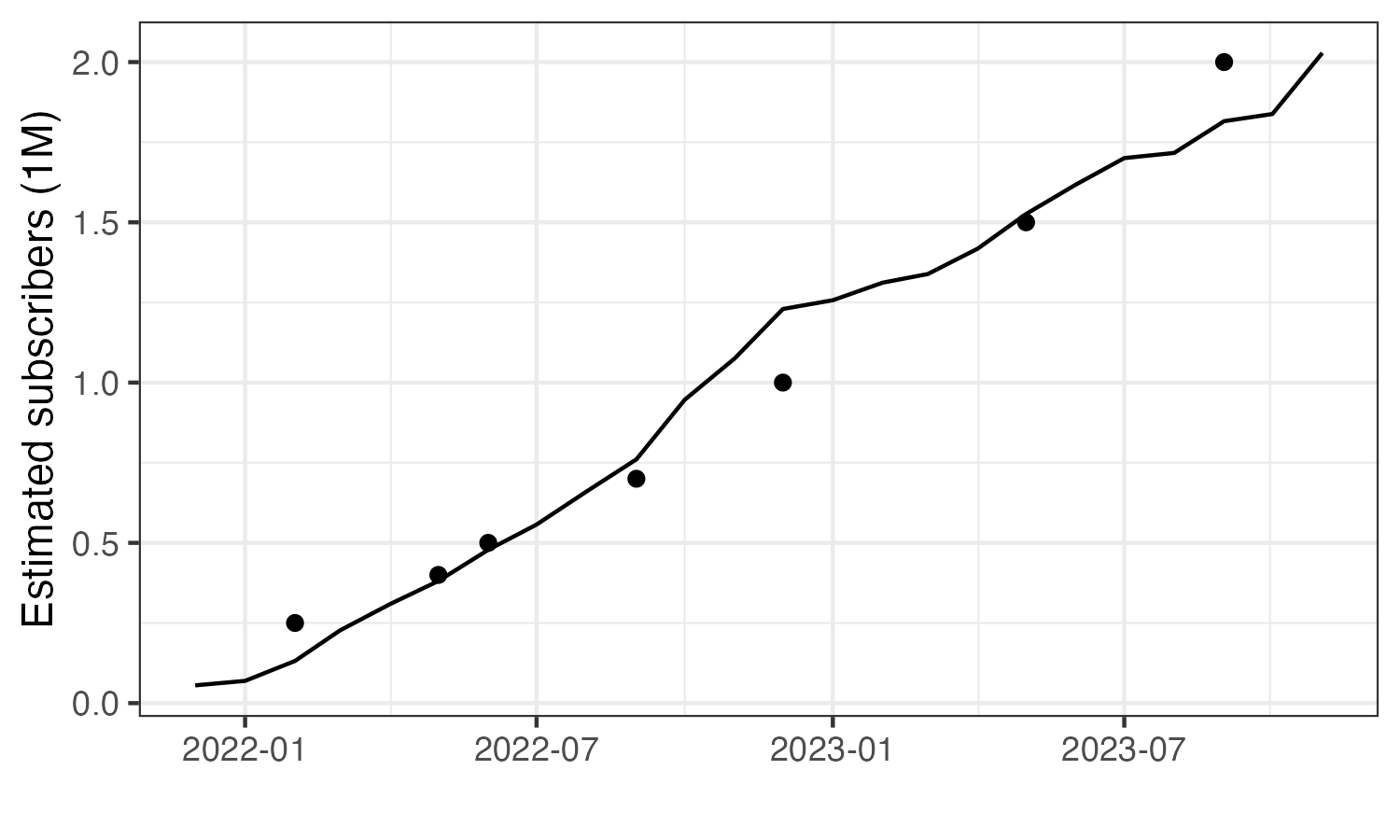}
\caption{Estimated weekly Starlink subscribers, global.}
\label{fig:global}
\end{subfigure}
\hfill
\begin{subfigure}[t]{0.48\textwidth}
\centering
\includegraphics[width=\textwidth]{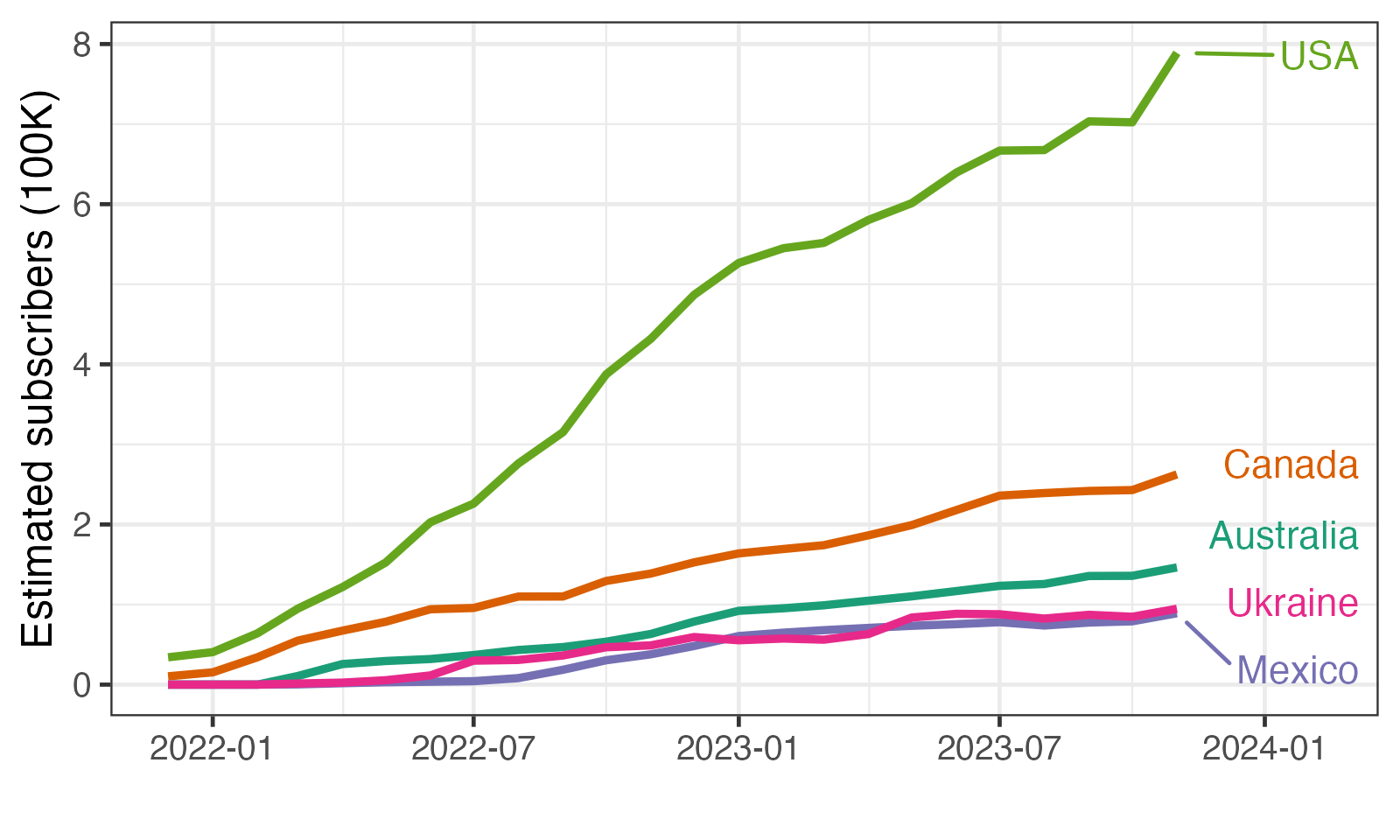}
\caption{Estimated weekly Starlink subscribers, top five countries.}
\label{fig:top5}
\end{subfigure}
\caption{Estimated weekly Starlink subscribers. Panel~(\subref{fig:global})
shows the global series (points are publicly reported counts; the line is the
DNS-traffic-derived estimate). Panel~(\subref{fig:top5}) shows the top five
countries by estimated November~2023 subscribers.}
\label{fig:subscribers}
\end{figure}

The processing to obtain weekly country-level subscriber estimates can be broken
into two steps. In the first step, the analysis used multiple publicly reported
counts of Starlink country-level subscribers, combined with the DNS traffic
data, to estimate the global weekly Starlink subscribers. This required applying
a statistical method called Ordinary Least Squares to scale DNS traffic to
publicly reported counts. The result of this step is shown in
Figure~\ref{fig:global}. The second step required combining the global weekly
Starlink subscriber estimates from the previous step with the normalized weekly
country-level traffic flows data and the country-level shares of Starlink web
traffic from Cloudflare. Figure~\ref{fig:top5} shows the resulting weekly
country-level Starlink subscriber estimates for the top five countries by
estimated Starlink subscriber counts in November~2023. Similarly, normalized
internet traffic from all sources was obtained from Cloudflare. Subtracting the
normalized Starlink traffic from the normalized internet traffic resulted in
estimates for the normalized traffic originating from non-Starlink sources.

GDP per capita data for 2022 was obtained from the World
Bank~\cite{worldbank} and used to group countries into income quartiles.
Starlink capacity on orbit was estimated from publicly available data on
Starlink deployments~\cite{gcat} and satellite
characteristics~\cite{krebs}, as well as publicly-available engineering
estimates of satellite data rates~\cite{rozenvasser}. A 10\% utilization factor
was applied to the total capacity estimate to obtain sold capacity, consistent
with public statements by Starlink engineers~\cite{kan}.

Traffic and capacity on orbit were aggregated from the weekly and daily
frequencies to the monthly frequency for statistical analysis. To study the
relationships between Starlink's monthly subscriber changes and capacity, the
following regression was estimated:
\begin{equation}
\log\left( Y_{it} \right) = \alpha \log(S_{t}) + \beta \log(B_{it})
+ \gamma \log(Y_{it - 1}) + \delta_{i} + \epsilon_{it},
\end{equation}
where $Y_{it}$ is estimated Starlink subscribers in country $i$ at time $t$,
$S_{t}$ is total Starlink capacity on orbit at time $t$, $B_{it}$ is estimated
background traffic in country $i$ at time $t$, $\delta_{i}$ is a set of
country-specific intercepts, and $\epsilon_{it}$ is the regression error. Since
the regression is written in logs, the coefficients $\alpha, \beta, \gamma$
measure average percentage changes in $Y_{it}$ for 1\% increases in $S_{t},
B_{it}, Y_{it-1}$ respectively, while $\delta_{i}$ measures the log subscribers
in each country in the initial period. The regression has an adjusted $R^{2}$ of
0.89, meaning 89\% of the variability in monthly country-level estimated
subscriber counts is explained by the included variables.

\section{Assumptions}

There are two main assumptions which may limit the applicability of, or findings
from, this method.

\paragraph{DNS traffic representativeness.} This method relies on DNS traffic
from a single provider. If Cloudflare DNS resolution is primarily used for
internet content that is not accessed by mega-constellation network users in
particular countries, then the demand estimates obtained by this method may be
systematically biased downwards in those countries and systematically biased
upwards in the other countries. This will not affect the global traffic
estimates but may cause distortions in the country-level traffic estimates. That
said, Cloudflare's DNS service is the most widely used in general, and use of
Cloudflare's DNS service among high-traffic websites is second only to Amazon's
DNS service~\cite{w3techs}, suggesting this is not likely to be a significant
issue. Future
improvements could source DNS traffic data from more providers.

\paragraph{Differences in average data use per subscriber.} This method assumes
average data use per subscriber is constant over time and countries. While this
is surely untrue, the question is how average data use per subscriber varies.
Secular time trends in average data use per subscriber will degrade
traffic-derived subscriber count estimates over time. However, neither
time-varying country-specific average data use per subscriber nor constant
country-specific differences in average data use per subscriber will bias
estimates of the relationships between mega-constellation consumer demand and
mega-constellation on-orbit capacity or background internet traffic from the
regression model described above. Such differences will be absorbed in the
variables controlling for previous-month traffic (in the case of time-varying
country-specific average data use per subscriber) and the country-specific
intercepts (in the case of constant country-specific differences in average data
use per subscriber). Future improvements could identify alternate sources for
average data use per subscriber over time and across countries.

\end{document}